\documentclass{article}  
\usepackage[left=2cm,right=2cm,top=2cm,bottom=2cm]{geometry}

\usepackage{amsmath,amsfonts,amssymb}
\usepackage{graphicx}
\usepackage[colorlinks=true, allcolors=blue]{hyperref}
\usepackage{subcaption}
\usepackage{mathtools}

\usepackage{tikz}
\usetikzlibrary{shapes,arrows}
\usetikzlibrary{decorations.pathreplacing}
\usetikzlibrary{decorations.pathmorphing, patterns,shapes}

\newcommand{\E}{\mathbb{E}}
\newcommand{\M}{\mathcal{M}}
\newcommand{\xh}{\hat{x}}
\newcommand{\yh}{\hat{y}}
\newcommand{\T}{\top}

\title{Optimal Intermittent Measurements for Tumor Tracking in X-ray Guided Radiotherapy}

\author{Antoine Aspeel$^a$, Damien Dasnoy$^a$, Rapha\"el M. Jungers$^a$ and Beno\^it Macq$^a$\\~\\
$^a$ICTEAM Institute, UCLouvain, Avenue Georges Lemaître 4-6, Louvain-la-Neuve, Belgium}

\begin{document} 
\maketitle


\begin{abstract}
In radiation therapy, tumor tracking is a challenging task that allows a better dose delivery. One practice is to acquire X-ray images in real-time during treatment, that are used to estimate the tumor location. These informations are used to predict the close future tumor trajectory. Kalman prediction is a classical approach for this task. The main drawback of X-ray acquisition is that it irradiates the patient, including its healthy tissues. In the classical Kalman framework, X-ray measurements are taken regularly, i.e. at a constant rate. In this paper, we propose a new approach which relaxes this constraint in order to take measurements when they are the most useful. Our aim is for a given budget of measurements to optimize the tracking process. This idea naturally brings to an optimal intermittent Kalman predictor for which measurement times are selected to minimize the mean squared prediction error over the complete fraction. This optimization problem can be solved directly when the respiratory model has been identified and the optimal sampling times can be computed at once. These optimal measurement times are obtained by solving a combinatorial optimization problem using a genetic algorithm. We created a test benchmark on trajectories validated on one patient. This new prediction method is compared to the regular Kalman predictor and a relative improvement of $9.8\%$ is observed on the root mean square position estimation error.
~
\\

\end{abstract}


\section{INTRODUCTION}

Mobile tumors are particularly challenging in radiation therapy. The motion creates uncertainty on tumor's position obliging the physicist to increase margins of the treatment that will irradiate healthy tissues. This effect is particularly important for particle therapy that is less robust to planning errors. A current response to this problem is to do image-guided radiotherapy. For particle therapy, it mainly consists on real-time X-ray acquisitions. In that case, X-ray images are acquired during the treatment to track the tumor in real-time at the cost of unwanted irradiation of the patient due to X-rays. There is a trade-off between more irradiations by imaging with smaller margins or less irradiations by imaging but with larger margins. For medical reasons (ALARA: As low as reasonably achievable), the measurement budget (amount of images) has to be restricted. These images feed a prediction model of the close future of tumor motion which is needed to compensate the delay between image acquisition and hardware reaction, including the computational time of the predictive algorithm. For now, these images are taken at constant rate. The purpose of this paper is (i) to relax this constraint by developing a predictive tumor tracking algorithm based on intermittent measurements; and (ii) to select the best moments to take the X-ray measurements in order to have the best tumor position estimation on the whole fraction.

The main idea is to optimally use each X-ray image allowing \textit{irregular sampling}, i.e. variable rate imaging. This additional degree of freedom permits a better use of each X-ray image and then, a better estimation of tumor's position without additional dose. More precisely, we consider that the number $N$ of X-ray images allowed during a fraction is fixed by the clinician. These $N$ measurements will be made when they are the more useful in order to reduce the error variance of the predicted tumor's location. The framework of Kalman filtering theory \cite{kalman1960new} has still be used by Sharp \textit{et al.} \cite{sharp2004prediction} in the context of tumor tracking in radiotherapy.

The rest of the paper is organized as follows: Section \ref{section:MM} develops how to use the method in clinical settings, the technical developments, an example on synthetic data and the kind of data; Section \ref{section:RD} presents and discuss of the prediction quality; finally Section \ref{section:conclusion} concludes with a summary of the results and further work.

\section{MATERIALS AND METHODS}\label{section:MM}
X-ray image acquisition is not directly suited for real-time because tumor localization and the measurement process encompass some delay. This delay imposes not only to estimate the current tumor position, but also to predict its future motion. Another physical restriction is the time between two X-ray acquisitions that can not be arbitrarily small. In this paper, these two delays are assumed to be equal. The time is discretized with a time step equal to that quantity. If the horizon prediction is not equal to the delay between X-ray images, the presented method can straitforwardly be adapted. In clinical workflow, X-ray images of the patient are acquired during the treatment. Following Sharp \textit{et al.} \cite{sharp2004prediction}, the first images are used to design the predictor and then, they are used for real-time tumor tracking.

In a first phase, these images are acquired regularly, i.e. at constant rate. These successive noisy measures $z(t)$ of the tumor location $y(t)$ are used to model the tumor motion. It is called \emph{system identification} (SI) and it is described in Subsection \ref{subsection:SI}.

Secondly, the model of the tumor motion is used to find the set $\M$ of $N$ optimal moments to take measurements with X-ray images. It corresponds to finding the \emph{mean squared estimator} (MSE) among all the \emph{intermittent Kalman predictors} (IKP). These measurement times are found by approximately solving a combinatorial optimization problem using a \emph{genetic algorithm} (GA). The modeling of this optimization problem is described in Subsections \ref{subsection:IKP} and \ref{subsection:optimalIKP}. Its resolution is in Subsection \ref{subsection:numeric}.

\begin{figure}[h]
\centering
\begin{tikzpicture}[scale=.25]
    \draw (0,0) -- (47,0);
    \draw (52,0) -- (60,0);
	\draw[decoration = {zigzag,segment length = 5mm, amplitude = 5pt},decorate] (47,0)--(52,0);
	

    \foreach \x in {0,...,20}
    		\draw (\x cm,15pt) -- (\x cm,-15pt);
    		
    	\foreach \x in {30,31,32, 36,37,38, 45,46, 53,54,55,56}
		\draw (\x cm,15pt) -- (\x cm,-15pt);
		
    \draw [thick,decorate,decoration={brace,amplitude=6pt,raise=0pt}] (0 cm,20pt) -- (19.9cm,20pt);
	\node[align=center] at (10cm,65pt) {Take regular measurements};
	
	\draw [thick,decorate,decoration={brace,amplitude=6pt,raise=0pt}] (20.1 cm,20pt) -- (29.9cm,20pt);
	\node[align=center] at (25cm,65pt) {SI and GA};
	
	\draw [thick,decorate,decoration={brace,amplitude=6pt,raise=0pt}] (30.1 cm,20pt) -- (60cm,20pt);
	\node[align=center] at (45cm,65pt) {IKP with $N$ intermittent measurements};
	
	\draw [thick,decorate,decoration={brace,amplitude=6pt,raise=0pt,mirror}] (30.1 cm,-20pt) -- (36.9cm,-20pt);
	\node[align=center] at (33.5cm,-65pt) {Warm up IKP};
	
	\draw [thick,decorate,decoration={brace,amplitude=6pt,raise=0pt,mirror}] (37.1 cm,-20pt) -- (60cm,-20pt);
	\node[align=center] at (48.5cm,-65pt) {Treatment};
  \end{tikzpicture}
  \caption{Time line of a fraction. Each phase is specified and vertical dashes represent measurement times. SI is system identification, GA is genetic algorithm and IKP is intermittent Kalman predictor.}
  \label{fig:timeLine}
\end{figure}
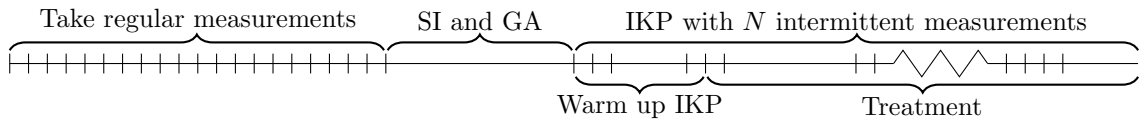

In a last phase, the tumor motion is predicted in real-time by the IKP based on measurements taken at the moments found by the GA. At the beginning of this phase, some time is left to warm up the predictor. The dose delivery begins when the predictor is warmed up. This procedure is represented on Figure \ref{fig:timeLine}.

\subsection{System identification}\label{subsection:SI}
We assume that the tumor position $y(t)$ can be modelled as a linear discrete-time process
\begin{equation}\label{modelxy}
\left\{
\begin{array}{rclr}
	x(t+1) &=& Ax(t)+b+Gw(t) &\text{(Internal states)}\\
	y(t)   &=& Cx(t)+d      &\text{(Tumor position)}
\end{array}
\right. t=0,\dots,T-1
\end{equation}
where $x(0)\sim\mathcal{N}(\bar{x}_0,\bar{P}_0)$, $w(t)\sim\mathcal{N}(0,Q)$ are mutually and time independent white Gaussian noises. $T$ is the duration of the treatment (in number of time steps). $x(t)$, $b$ and $w(t)$ are $n\times 1$ vectors; $A$ is a $n\times n$ matrix; $G$ has compatible dimensions; $y(t)$ and $d$ are $m\times 1$ vectors and $C$ is a $m\times n$ matrix. In practice, $m$ is 2 or 3 for 2- and 3-dimensional tracking, respectively. The number of internal states $n$ is an hyper-parameter to determine (it is not patient-specific). The system identification problem is to model the tumor position $y(t)$ on the basis of noisy measurement $z(t)$. The measurement budget is fixed to $N$, therefore the set of measurement times is $\M=\{m_0,\dots,m_{N-1}\}\subset\{0,\dots,T-1\}$. It gives the measurement equation $z(t) = y(t) + v(t)$ for any $t\in\M$ in which the white Gaussian noise $v(t)\sim\mathcal{N}(0,R)$ is time independent and independent of $x(0)$ and $w(\tau)$ for any $\tau$. Each $z(t)$ is a $m\times 1$ vector. The previous measurement equation and equations (\ref{modelxy}) can be rewritten to fit the Kalman formalism,
\begin{equation}\label{modelxyz}
\left\{
\begin{array}{rclcrcl}
	x(t+1)                &=& Ax(t)+b+Gw(t)  & & t &=  & 0,\dots,T-1 \\
	z(t)-d   &=& Cx(t) + v(t)  & & t &\in& \M
\end{array}
\right.
\end{equation}
where $b$ can be considered as a constant command. Note that during the first phase (see Figure \ref{fig:timeLine}), X-rays are acquired regularly and so, during system identification, the second equation has to be considered for all $t$.

The purpose of these equations is to model the tumor motion, thus the parameters of this system are identified from regularly spaced measurements at the beginning of the fraction, they are $A$, $b$, $Q$, $d$, $C$, $R$, $\bar{x}_0$ and $\bar{P}_0$. This system identification is done using the maximum likelihood estimation with the expectation maximization algorithm of Digalakis \textit{et al.} \cite{digalakis1993ml} as suggested by Sharp \textit{et al.} \cite{sharp2004prediction} in the context of tumor tracking.

\subsection{Intermittent Kalman predictor}\label{subsection:IKP}
Let's define\footnote{The symbol $\coloneqq$ is used for definitions.} $\xh(t|t-1)\coloneqq\E[x(t)|z(m_i), m_i< t]$ as the best \emph{a priori} MSE of $x(t)$ and $\xh(t|t)\coloneqq\E[x(t)|z(m_i), m_i\leq t]$ as the best \emph{a posteriori} MSE of $x(t)$, where the $m_i$ are in $\M$. In addition, one can define the a priori and the a posteriori covariance matrices respectively as $P(t|t-1) \coloneqq \E[(x(t)-\xh(t|t-1))(x(t)-\xh(t|t-1))^\T]$ and $P(t|t) \coloneqq \E[(x(t)-\xh(t|t))(x(t)-\xh(t|t))^\T]$. The classical Kalman filtering theory \cite{kalman1960new} states how to recursively update that four quantities in the case where a measurement is taken at each time step, i.e. when $N=T+1$. In addition, define $\hat{y}(t)\coloneqq C\hat{x}(t)+d$, an unbiased estimator of $y(t)$.

Inspired by Sinopoli \textit{et al.} \cite{sinopoli2004kalman}, we consider the intermittent case by replacing $R$ by $\lim_{\mu\rightarrow\infty}\mu I$ when no measurement is available (where $I$ is the $m\times m$ identity matrix). This is driven by the fact that an infinite noise measurement is equivalent to no measurement. It follows that when no measurement is available, each a posteriori quantity is simply the a priori one, what could have been expected. Then the equations of the IKP are \textbf{time update} equations
\begin{equation}\tag{3, 4, 5}\addtocounter{equation}{3}
\begin{array}{cccc}
P(t+1|t) = AP(t|t)A^\T + GQG^\T, &
\xh(t+1|t) = A\xh(t|t) + b & \text{and} &
\yh(t+1|t) = C\xh(t+1|t) + d
\end{array}
\end{equation}
for $t=0,\dots,T-1$. In addition, \textbf{measure update} equations are
\begin{equation}\label{MU_P}
P(t|t) = \left\{
	\begin{array}{ll}
		\left[P(t|t-1)^{-1}+C^{\T}R^{-1}C\right]^{-1} &\text{if}\ t\in\M \\
		P(t|t-1) &\text{else},
\end{array}\right.
\end{equation}
\begin{equation}\label{MU_x}
\xh(t|t) = \left\{
	\begin{array}{ll}
		\xh(t|t-1) + P(t|t)C^{\T}R^{-1}\left[z(t)-d-C\xh(t|t-1)\right] &\text{if}\ t\in\M \\
		\xh(t|t-1) &\text{else},
	\end{array}\right.
\end{equation}
\begin{equation}
\yh(t|t) = C\xh(t|t)+d
\end{equation}
for $t=0,\dots,T$. The \textbf{initialization} of these recurrence equations are
\begin{equation}\label{initialization}
\begin{array}{cccc}
P(0|-1)=\bar{P}_0, & \xh(0|-1)=\bar{x}_0 & \text{and} & \yh(0|-1)=C\bar{x}_0 + d.
\end{array}
\end{equation}
The IKP is summarized by equations (3) to (\ref{initialization}).

\subsection{Optimal intermittent Kalman predictor}\label{subsection:optimalIKP}
The optimal IKP is defined as the IKP that minimizes the error variance. The set of measurement times $\M$ is selected to minimize that quantity.

The covariance matrix of the a priori error $y(t)-\yh(t|t-1)$ can be written
\begin{align*}
S(t|t-1) &\coloneqq \E\left[ (y(t)-\yh(t|t-1)) (y(t)-\yh(t|t-1))^\T\right]\\
&\ =\E\left[ (Cx(t)-C\xh(t|t-1)) (Cx(t)-C\xh(t|t-1))^\T \right]\\
&\ =C\E\left[ (x(t)-\xh(t|t-1)) (x(t)-\xh(t|t-1))^\T \right]C^\T\\
&\ =CP(t|t-1)C^\T
\end{align*}
and similarly for the corresponding a posteriori covariance matrix
$$
S(t|t) \coloneqq \E\left[ (y(t)-\yh(t|t)) (y(t)-\yh(t|t))^\T\right]=CP(t|t)C^\T.
$$
The variance of the prediction error $y(t)-y(t|t-1)$ can be written
$$
\E\left[\| y(t)-\yh(t|t-1) \|^2\right] = \text{Tr}[S(t|t-1)] = \text{Tr}[CP(t|t-1)C^\T ]
$$
where $\text{Tr}[\cdot]$ is the trace operator. 

The set of measurement times $\M$ is chosen to minimize this quantity over the complete treatment, i.e. from the end of the warm up $T_0+1$ to the end of the fraction $T$. In other words, the \emph{optimal intermittent Kalman predictor} is an IKP for which the set of measurements $\M$ is the solution of the following optimization problem
\begin{equation}\label{optimalIKP}
\min_{\M\subset \{0,\dots,T-1\}} \sum_{t=T_0+1}^{T}\text{Tr}[CP(t|t-1)C^\T] \ 
\text{subject to}\ 
|\M| = N,\ 
\text{equations (3) and (\ref{MU_P}) with}\ 
P(0|-1)=\bar{P}_0
\end{equation}
where constraints (3) and (\ref{MU_P}) are considered for $t=0,\dots,T-1$ and $t=0,\dots,T$, respectively.

One can observe that equations governing covariance matrices, i.e, equations (3) and (\ref{MU_P}), are independent of the measurements $z(t)$. It means that the optimization problem (\ref{optimalIKP}) can be solved before measurements are made and ignoring equations involving $\xh$ and $\yh$.

\subsection{Numerical computation of the optimal measurement times}\label{subsection:numeric}
This subsection briefly presents an algorithm to solve problem (\ref{optimalIKP}). It is a combinatorial optimization problem with $\dfrac{T!}{N! (T-N)!}$ possibilities for $\M$ which is computationally intractable. For large values of $T$, an exact algorithm to solve it runs out of time, even with a careful implementation. A fast search algorithm is developed at the price of a suboptimal solution. It is a standard genetic algorithm \cite{mitchell1998introduction} where each candidate solution is a vector $\begin{pmatrix}m_0,\cdots,m_{N-1}\end{pmatrix}$ which represents one choice of $\M$. After mutations and crossovers, duplicates can appear, i.e. $m_i=m_j$ even if $i\neq j$. In that case, same values are replaced by uniformly picking a time in $\{0,\dots,T-1\}\backslash\{ m_0,\cdots,m_{N-1} \}$. This procedure ensures that $m_i\neq m_j$ if $i\neq j$ and speed up the convergence of the algorithm.

\subsection{Example on synthetic data}
In this section, we compare the optimal IKP to the classic Kalman predictor on simulated data. A toy model of one dimensional motion simulates the tumor position. We consider a mass-spring equation subject to a random force $w(t)$
$$
\dfrac{d^2y}{dt^2}(t) = - \dfrac{k}{m} y(t) + \dfrac{1}{m}w(t)
$$
where $y(t)\in\mathbb{R}$ is the spring elongation at time $t$, $k$ is the stiffness of the spring and $m$ is the mass. $w(t)$ is a white Gaussian noise of power spectral density $Q$.  It can be written in vector form as
$$
\dfrac{dx}{dt}(t)=Ax(t) + Gw(t)
$$
where $x(t) = \begin{pmatrix} y(t) & \frac{dy}{dt}(t)\end{pmatrix}^\T$,  $A = \begin{pmatrix}0&1\\-k/m&0\end{pmatrix}$ and $G = \begin{pmatrix} 0 & 1/m \end{pmatrix}^\T$.

This equation can be discretized at period $\delta$ by defining $x^n\coloneqq x(\delta n)$ for all $n\in\mathbb{N}$. If $\delta$ is sufficiently small, it can be approximated as\footnote{One can observe that the following recurrence equation is unstable. This does not affect the validity of our purpose. Indeed, the mathematical framework of this paper includes unstable dynamics.}
$$
x^{n+1} = (I+A\delta)x^n + Gw^n
$$
with $w^n \sim\mathcal{N}(0,Q\delta )$.

A noisy measurement $z^n$ of the spring elongation $y^n = y(\delta n)$ gives the following equation
$$
z^n =y^n+v^n = \begin{pmatrix}1&0\end{pmatrix}X^n + v^n
$$
where $v^n\sim\mathcal{N}(0,R)$ is independent of $w^p$ for all $p$.

In this numerical example, we fix $m = 0.0015[kg]$ ($1.5[g]$) which is a reasonable mass for a tumor. In addition, we want a period $\bar{T}=2\pi\sqrt{m/k}=8[sec]$ that approximately corresponds to one breathing cycle then, we fix $k = 4\pi^2m/\bar{T}^2 = 0.000825[N/m]$. The discretization time is fixed to $\delta = 0.01[sec]$. The duration is fixed to $T = 6000$ time steps, i.e. $60[sec]$ and the number of measurements to $N = 20$. Finally, $Q=0.05[mm^2]$ and $R =0.05[mm^2]$. No warm up is considered, i.e. $T_0=0$.

\begin{figure}[!t]
\centering
\includegraphics[scale=.8]{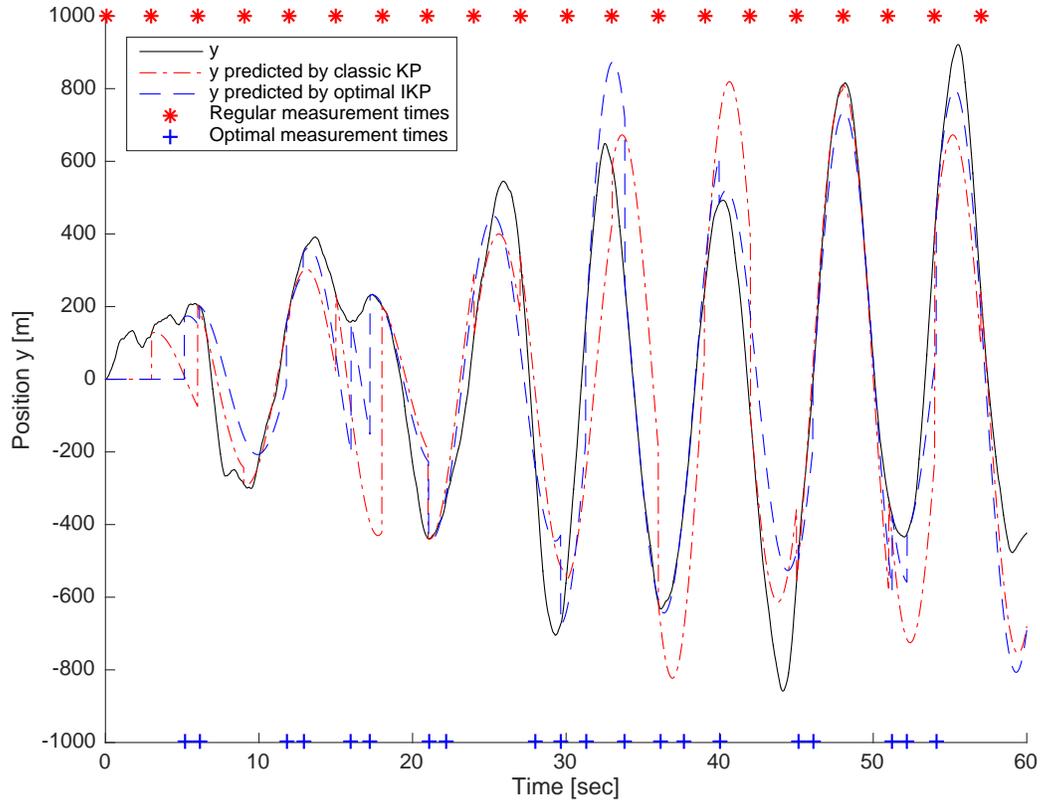}
\caption{Comparison of the predictions of the classical Kalman predictor (KP) and the optimal intermittent Kalman predictor (IKP) on simulated data. Measurement times are also indicated for both cases.}
\label{fig:simulatedData_x}
\end{figure}

\begin{figure}[!t]
\centering
\includegraphics[scale=.5]{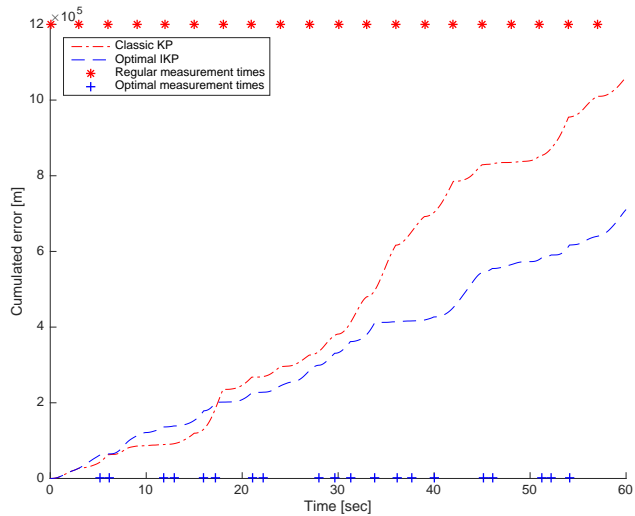}
\caption{Comparison of the cumulated errors (in absolute values) of the classical Kalman predictor (KP) and the optimal intermittent Kalman predictor (IKP) on simulated data. Measurement times are also indicated for both cases.}
\label{fig:simulatedData_cumErr}
\end{figure}

The best objective value of (\ref{optimalIKP}) found by the genetic algorithm is $476\cdot 10^6$ against $608\cdot 10^6$ for the regular measurements case. The predictions with the regular Kalman predictor and with the optimal IKP are depicted on figure \ref{fig:simulatedData_x}. On that realisation, the mean prediction error is about $118[m]$ with the optimal IKP against $176[m]$ for the regular Kalman predictor. It is an improvement of $33\%$. Figure \ref{fig:simulatedData_cumErr} presents the cumulated prediction error. One can observe that after a short beginning, the IKP cumulated error stays smaller than the cumulated error of the classic Kalman predictor.

\subsection{Data description}
\begin{figure}[!t]
\centering
\includegraphics[scale=.5]{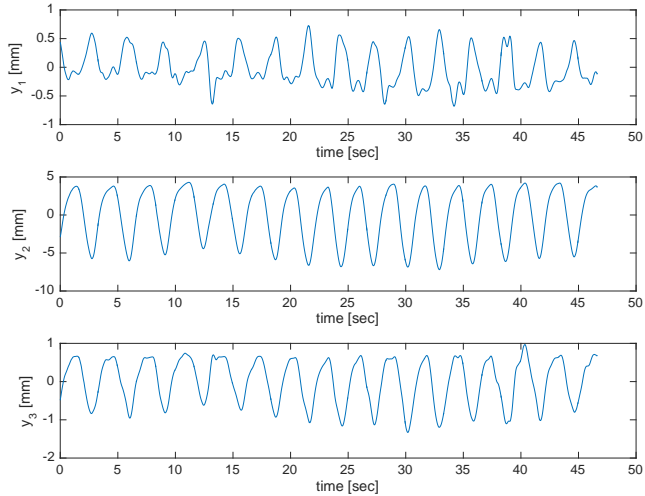}
\caption{The tumor position $y(t)=\left(y_1(t)\ \ y_2(t)\ \ y_3(t)\right)^\T$ with respect to time for the studied patient. Figure inspired from figure 1 of Bukovsky \textit{et al.} \cite{bukovsky2015fast}.}
\label{fig:sgroundTruth}
\end{figure}

The tumor position $y(t)$ is obtained from the figure 1 in Bukovsky \textit{et al.} \cite{bukovsky2015fast} and are depicted in figure \ref{fig:sgroundTruth}. $y_1$, $y_2$ and $y_3$ are the position on the lateral, cephalocaudal and anteroposterior axes, respectively.

The trajectory is considered as the ground truth $y(t)\in\mathbb{R}^3$. A noisy trajectory $z(t)$ is simulated from this ground truth to imitate a X-ray acquisition followed by a tracking software, i.e. $z(t) = y(t) + v(t)$ with $v(t)\sim\mathcal{N}(0,\sigma^2I)$. The parameter $\sigma^2$ quantifies the quality of the observation process, the more the measurement procedure is precise, the more $\sigma^2$ is small. For example, a high $\sigma^2$ models a low-dose imaging. The prediction algorithm uses $z$ as noisy measurements of the tumor and computes estimate $\yh(t|t-1)$ of the ground truth $y(t)$.

Real time between two discrete time steps is $1/30$ seconds. The training duration is 600 time steps (20 seconds), we fix $T=800$ (26.7 seconds), and warm up takes 300 time steps (10 seconds). In the next section, we compare our algorithm to other methods.

\section{RESULTS AND DISCUSSION}\label{section:RD}
Hyper-parameters have been tuned on simulated data. Our method (IKP) is compared to three other methods: \emph{regular measurements without prediction} (RMWP), i.e. the tumor is assumed motionless between measurements; \emph{intermittent measurements without prediction} (IMWP), i.e measurements are taken at the same time than IKP but tumor is assumed motionless between measurements; and \emph{regular Kalman predictor} (RKP), i.e. with regularly spaced measurement times. Root-mean square prediction errors are presented in Table \ref{tab:results} for these four methods. Values of $\sigma^2\in\{1,\  4,\ 25,\ 100\}$ are tested, it corresponds to a standard deviation of $1,\ 2,\ 5,\ 10\ [mm]$, which is a plausible range of measurement error for practical applications. In addition, the comparison is done for different measurement budget $N$ ($N/T$ and mean acquisition rate are printed for more meaning).

\begin{table}[h!]
\centering
\begin{tabular}{|c|c||c|c|c|c|}
\hline 
\multicolumn{2}{|r||}{$\sigma^2$ [$mm^2$]} & 1 & 4 & 25 & 100 \\
\hline
$N/T$ (mean rate)& Method & & & & \\
\hline 
\hline 
     & RMWP & 2.48 & 3.99 & 9.14 & 18.01 \\
0.1  & IMWP & 2.22 & 3.93 & 9.05 & 17.90 \\
(3 [$Hz$])    & RKP  & 1.29 & 1.71 & 3.51 & 3.78 \\
     & IKP  & \textbf{1.11} & \textbf{1.50} & \textbf{3.14} & \textbf{3.64} \\
\hline 
     & RMWP & 1.99 & 3.67 & 8.96 & 17.85 \\
0.2  & IMWP & 2.00 & 3.59 & 8.82 & 17.53 \\
(6 [$Hz$])    & RKP  & 1.02 & 1.42 & 3.04 & 3.61 \\
     & IKP  & \textbf{0.98} &\textbf{ 1.15} & \textbf{2.86} & \textbf{3.54} \\
\hline 
     & RMWP & 1.88 & 3.56 & 8.77 & 17.48 \\
0.3  & IMWP & 1.87 & 3.54 & 8.79 & 17.54 \\
(9 [$Hz$])    & RKP  & 0.95 & 1.31 & 2.77 & 3.47 \\
     & IKP  & \textbf{0.81} & \textbf{1.20} & \textbf{2.47} & \textbf{3.31} \\
\hline 
     & RMWP & 1.84 & 3.56 & 8.82 & 17.62 \\
0.4  & IMWP & 1.84 & 3.52 & 8.83 & 17.64 \\
(12 [$Hz$])    & RKP  & 0.84 & 1.24 & 2.63 & 3.43 \\
     & IKP  & \textbf{0.76} & \textbf{1.09} & \textbf{2.27} & \textbf{3.14} \\
\hline 
     & RMWP & 1.79 & 3.51 & 8.72 & 17.42 \\
0.5  & IMWP & 1.82 & 3.54 & 8.78 & 17.52 \\
(15 [$Hz$])  & RKP  & 0.83 & 1.21 & 2.59 & 3.44 \\
     & IKP  & \textbf{0.73} & \textbf{1.06} & \textbf{2.29} & \textbf{3.23} \\
\hline 
\end{tabular}
\caption{Root-mean square prediction errors [$mm$] for different predictors, different $\sigma^2$ and different $N/T$ (also written $N/T\times 30Hz$, the mean acquisition rate). Compared methods are RMWP: Regular measurements without prediction; IMWP: Intermittent measurements without prediction; RKP: Regular Kalman predictor; and our method, IKP: Intermitent Kalman predictor. Bold characters highlight the best predictor in each situation.}
\label{tab:results}
\end{table}
Table \ref{tab:results} shows that our method (IKP) outperforms the three other predictors in all cases. Globally IMWP gives similar results than RMWP, indicating that the measurement times computed for IKP are well-suited only for Kalman predictions. The two Kalman approaches (RKP and IKP) are highly more precise than when there is no predictor (RMWP and IMWP). For the values presented in that table, the mean relative improvement of IKP on RKP is about\footnote{It is $(\text{RMS}_\text{RKP} - \text{RMS}_\text{IKP})/\text{RMS}_\text{RKP}$ averaged on the values of Table \ref{tab:results}, where RMS means root-mean square error.} $9.8\%$. If some improvements look marginal, note that a small improvement in tumor localization permits to reduce the margins of the planning target volume. In addition, the tumor being a three-dimensional object, when the margin radius $r$ decreases, the planning target volume decreases in $r^3$ which amplifies the benefit, specifically for precise dose delivery by particle therapy.

Nevertheless, a demonstration on only one patient has to be considered as a proof of concept but is not sufficient for clinical application.

\section{CONCLUSION}\label{section:conclusion}
A tumor tracking method based on an optimal choice of measurement times has been developed. The benefit of using well-chosen intermittent measurement times in the case of radiotherapy guided by X-ray images has been shown. This improvement is particularly interesting in the case of particle therapy because of the reduction errors on tumor localization has significant effects on the margin reduction (sparing healthy tissues).

The presented method is an optimal intermittent Kalman predictor. Its development has required the formulation of a combinatorial optimization problem to find the best moments to take a fixed number of measurements (X-ray images). It lays the foundations of a mathematical framework for optimal intermittent Kalman prediction and filtering. Then, a genetic algorithm has been implemented to solve approximately this combinatorial problem with a limited complexity. The first experiments that have been presented show that this new method seems to outperform the classic Kalman approach, i.e. with regular measurement times. However, we prefer to be cautious about our conclusions because the evidences that have been presented are based on only one patient. Deeper tests are needed to definitely assess the performance of this method on clinical framework, even if the proof of concept seems to be encouraging. Further work could also study a sliding version of the intermittent Kalman predictor as suggested by Riaz \textit{et al.} \cite{riaz2009predicting}, it means that the system identification is continuously updated when new measurements are obtained.

\section*{ACKNOWLEDGMENTS}
A.A. is supported by the Walloon Region, its grant is RW-DGO6-Biowin-Bidmed.
R.J. is a FNRS Research Associate. He is supported by the Walloon Region and the Innoviris Foundation.

\bibliographystyle{spiebib} 
\bibliography{report} 

\end{document}